# Does Task Complexity Moderate the Benefits of Liveness? A Controlled Experiment

Patrick Rein[a], Stefan Ramson[a], Tom Beckmann[a], and Robert Hirschfeld[a]

a   Hasso Plattner Institute, University of Potsdam, Germany

**Abstract**   Live programming features can be found in a range of programming environments, from individual prototypes to widely used environments.

While liveness is generally considered a useful property, there is little empirical evidence on when and how liveness can be beneficial. Even though there are few experimental studies, their results are largely inconclusive.

We reviewed existing experiments and related studies to gather a collection of potential effects of liveness and moderating factors. Based on this collection, we concluded that *task complexity* and *prior experience addressing liveness* are potentially essential factors neglected in previous experiments. To fill this gap, we devised and conducted a controlled experiment ($N = 37$) testing the hypothesis that task complexity moderates the effects of live introspection tools on participants' debugging efficiency, given participants with prior experience with liveness.

Our results do not support the hypothesis that task complexity moderates the effect of live introspection tools. This non-significant moderation effect might result from the low number of participants, as the data suggests a moderation effect. The results also show that in our experiment setting, live introspection tools significantly reduced the time participants took to debug the tasks.

For researchers interested in the effects of liveness, our findings suggest that studies on liveness should make conscious choices about task complexity and participants' prior experience with liveness. For designers of programming environments, the results of our experiment are a step toward understanding when and how programming tools should be live to become more helpful to programmers.

## The Art, Science, and Engineering of Programming

**Does Task Complexity Moderate the Benefits of Liveness? A Controlled Experiment**

# 1 Introduction

Several research communities have worked on an experience of liveness in programming in various application domains, programming environments, and programming languages. Numerous live programming tools and environments have been created, and even widespread programming systems, such as Microsoft Excel and Jupyter Notebooks, support liveness to some degree.

Liveness as "an impression of changing a program while it is running" [51] is generally regarded as a beneficial feature. Designers of programming environments argue that liveness can improve program comprehension, domain exploration or engagement [26, 51, 64]. For example, one theory claims that when professional programmers are debugging a program through liveness, "the important relationships are made manifest" [64]. Another theory claims that liveness could foster exploration and engagement of children in learning environments [26]. In general, user studies have repeatedly shown that programmers regard liveness as a desirable feature [10, 34, 38], which fits into a larger trend hinting that programmers favor short feedback loops in many kinds of programming tools [42, 63].

At the same time, there are only a few experimental studies on the benefits of liveness, and their results are either inconclusive or only show that liveness is not beneficial in their experimental setting [27, 34, 65]. For instance, one study investigated the impact of liveness on programmers' debugging accuracy in spreadsheet environments and only found an effect in one of two tasks [65]. Another study investigated the impact of liveness on novices' programming effectiveness and also found no effect in general [27]. We argue that the factors determining when programmers benefit from liveness are more complex than previously assumed in experiments.

As a first contribution, we review existing empirical accounts of liveness to describe the potential effects of liveness and factors that may influence this effect. Further, based on this collection of factors, we conclude that two factors have been neglected in previous studies: task complexity and participants' experience with liveness. As the main contribution, we describe the setup and results of a controlled experiment ($N = 37$) testing the hypothesis: *task complexity moderates the impact of live introspection tools on debugging performance for programmers experienced with live programming tools*. The results are inconclusive regarding the moderation effect but show that liveness did improve debugging efficiency in our experiment setting.

**Outline** In the remainder of the paper, we first describe previous studies investigating the effects of liveness and related empirical insights from human-computer interaction (HCI) in Section 2. We summarize the results in a collection of potential effects of liveness and factors influencing them. We then describe the general design of the controlled experiment in Section 3 including the independent and dependent variables, the application scenario, the task creation process, and the participant selection. We describe the results in Section 4 and discuss them in Section 5.

## 2  Background on Related Studies on Liveness and Feedback Loops

Tools incorporating liveness are often motivated through theories that predict that live programming should be beneficial [26, 44, 64]. Taken together, these theories claim a variety of benefits of liveness in a wide spectrum of programming settings. While there are some evaluations of individual tools and systems, for instance [16, 22, 23], those broad claims have only little experimental backing from dedicated, fixed-setup, empirical studies on the effects of liveness (see Figure 1). Despite the variety of claims, only a small number of studies have been conducted that investigate the impact of liveness in particular. Further, the few existing studies indicate that liveness might not be beneficial in all of the settings proclaimed by theory [27, 28, 65].

From these studies on liveness and related studies on feedback loops in general, we derive a set of aspects that liveness may influence and a set of factors that may affect the impact of liveness.

### 2.1  Perspective on Liveness

To contextualize the following studies, we briefly discuss different perspectives on liveness. There are at least three different perspectives on liveness in programming environments represented by the terms *exploratory programming*, *live programming*, and *live coding* [51].

**Exploratory Programming**  The first perspective is that of exploratory programming environments [56, 62]. While the term *exploratory programming* refers to a general approach to programming, *exploratory programming environments* refers to a particular group of environments that focus on changing a running system, including both its code and state.

**Live Programming**  The second perspective is live programming [11, 26, 37], which aims to improve code creation by providing immediate feedback on the dynamic behavior of code. The immediacy of feedback is achieved by providing *automatic feedback* without requiring programmers to manually request it.

**Live Coding**  The third perspective is live coding [7, 13], which employs liveness in art performances based on programs that often produce acoustic or visual effects. Programming thereby becomes the *real-time adjustment* of running programs producing a desired effect.

### 2.2  Previous Experiments on Liveness

In the following, we will describe the published empirical studies on liveness in programming environments we could locate and their settings and observed effects. Due to the small number of existing controlled experiments, we describe all experiments we located that investigate liveness in general, despite our experiment focusing on the liveness provided by exploratory programming environments. The list of studies is based on a previous literature study on liveness [51], which we extended with selected studies that are newer or are not controlled experiments but contribute relevant

**Does Task Complexity Moderate the Benefits of Liveness? A Controlled Experiment**

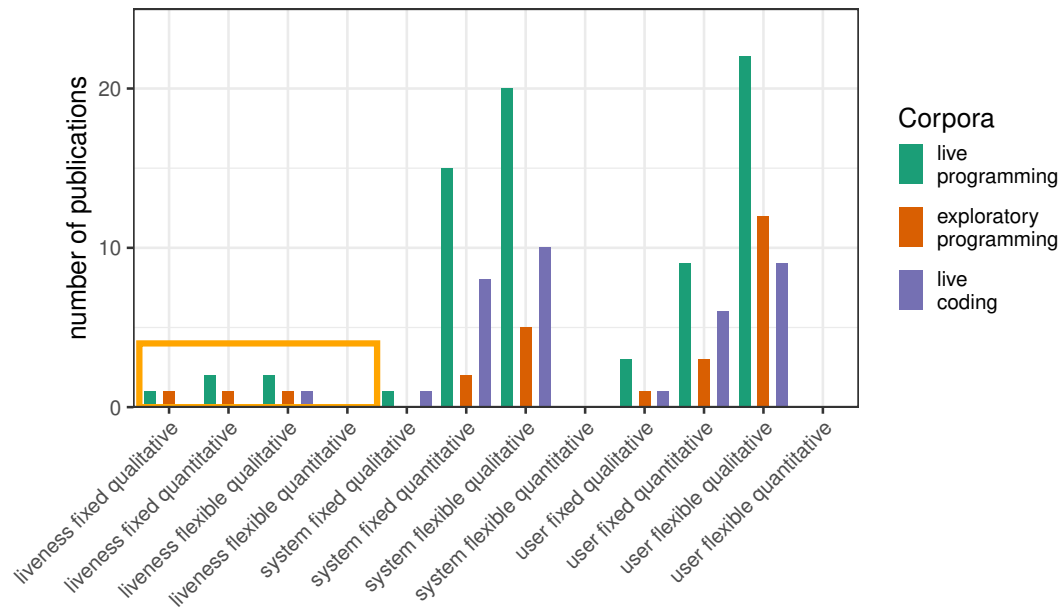

■ **Figure 1** A bar chart showing the number of types of empirical studies in each corpus (exploratory programming, live programming, live coding) [51]. The studies are characterized according to the studied subject (liveness, system, user), whether the study procedure was fixed beforehand or modified in the course of the study (fixed, flexible) [53], and what kind of data was gathered (quantitative, qualitative). Notably, there is a general lack of empirical studies having liveness itself as the studied subject (bars in the highlighted area).

insights. Notably, we only selected studies that investigated liveness of some form and excluded studies that evaluated a specific tool or system. For each experiment or study, we will describe the controlled and measured variables and the variables mentioned in future work sections.

### 2.2.1 An Experimental Study of the Impact of Visual Semantic Feedback on Novice Programming

This controlled experiment investigated the effects of feedback modes on the novice programmers' programming effectiveness in a new live programming environment combined with a live algorithm visualization tool and a newly created programming language [27, 28].

The authors selected 57 university students from an introductory programming course, keeping the programming and liveness experience constant at the novice level. All tasks were of low general complexity, such as summing up an array. They varied the *feedback mode* between no feedback, self-selected feedback by pressing a button, and automatic feedback on changes to the code.

Overall, the experiment did not show any significant effect as the range of the results was too broad. This might indicate that liveness is not generally beneficial for absolute novice programmers, which is in line with studies showing that they

1:4

already struggle with fundamental programming concepts [52]. The study identified two variables as future work: readiness for feedback and feedback delay.

### 2.2.2 Does Continuous Visual Feedback Aid Debugging in Direct-Manipulation Programming Systems?

This controlled experiment investigated the effects of feedback modes on the debugging effectiveness of experienced programmers in a new spreadsheet live programming environment [14, 65].

The authors selected 29 participants from a senior-level operating systems course. Participants had considerable programming experience. Their experience with liveness is not reported in detail beyond reported exposure to spreadsheets and LISP. Each participant got two tasks of medium complexity. One task was debugging the control logic for a seven-segment display, and the other was debugging a formula to check a hash. The seven-segment task involved seven defects, and the hash task involved five defects. There was a time limit of 15 minutes per task. The *feedback modes* varied between automatic and self-selected feedback, which was also delayed for 90 seconds.

The experiment did not find any difference in the overall debugging accuracy. However, there was a significant difference for each task. For the mathematical task, the automatic feedback group performed better, while for the seven-segment display task, it performed worse. Further, participants in the automatic feedback group performed more changes. Finally, in a post-study questionnaire, the automatic feedback group reported higher confidence in their understanding of the problems. Based on the different results for the two tasks, the authors mention task domain and *task complexity* as relevant moderator variables for future work. Further, they mentioned participants' performance on the tasks as another future control variable.

### 2.2.3 How Live Coding Affects Developers' Coding Behavior

This controlled experiment investigated the effect of feedback modes on the code creation effectiveness of experienced programmers in a new live programming environment for JavaScript [34].

The authors selected ten participants, some of them students, with at least four and a half years of programming experience. The participants' experience with liveness was not discussed. Each participant got three tasks of medium complexity. While the domain was complex, for example using a previously unknown parser library, the task only required to implement one function. The feedback modes were varied between no feedback and automatic feedback. At the same time the environment allowed access to introspection tools through the Chrome debugger.

The experiment found no difference in the number of total defects introduced. However, in the automatic feedback group defects were repaired more quickly. The overall task completion time was the same for both groups. The development strategy was significantly different between the groups, with the automatic feedback group using an interleaved and the manual feedback group using a sequential strategy of changing code and examining run-time behavior. All participants stated that through automatic feedback they became more confident that their code was correct. No additional variables were given as future work.

#### 2.2.4 The Road to Live Programming: Insights From the Practice

This study does not describe a controlled experiment, but, besides other investigations, a fixed-setup, quantitative study on the usage patterns of tools showing dynamic or static information on the system [35].

During the quantitative study, the authors observed 11 participants while programming in an exploratory-style live programming environment. The participants all reported more than three years of programming experience. Most of them also reported more than one year of experience with live programming. Only two participants reported half a year of experience with liveness.

The general result of this study is that tools for inspecting run-time state are used frequently in exploratory-style live programming environments. While the difference between novice and experienced live programmers was not part of the study, the article notes that on some occasions, the programmers with little live experience did not think of run-time introspection tools as a means to explore the system.

#### 2.2.5 Edit-Run Behavior in Programming and Debugging

This study also does not describe a controlled experiment but an exploratory, observational study on edit-run cycles during programming and debugging phases [1]. The motivation for the study explicitly mentions that the results should inform future designs of live programming tools by better understanding the current state of edit-run cycles.

The authors used 28 hours of coded recordings of programming sessions from 11 professional programmers. They then automatically determined the programming and debugging phases and determined edit-run cycles within these phases.

They found that programmers use considerably more edit-run cycles during debugging than editing. While the program runs also include the debugger, this result is interesting, as it highlights how heavily programmers rely on dynamic information during debugging. During editing, programmers most often edited a single file and then ran the program once, presumably to check the result of their edit.

### 2.3 Other Relevant Variables as Indicated by Related Research Fields

Aspects such as short feedback loops or potential effects, such as improved debugging effectiveness, have been studied extensively outside the context of liveness. In turn, the results of these studies hint at relevant variables for studies in the live programming context. In particular, we look into the impact of programming experience on the usage of complex programming tools and mechanisms and the variables moderating the effects of short feedback loops.

#### 2.3.1 Experienced and Novice Programmers

Liveness and the resulting tools and workflows can be considered an advanced mechanism of programming systems. Research on the impact of programming experience shows that novices use advanced mechanisms, such as complex programming tools or language features, less often and less effectively [25, 52].

So far, there are no dedicated studies on the impact of programming experience on the effects of liveness. However, related studies indicate that it might influence the effectiveness of using a live system. One survey of studies on programming pedagogy found that novices mostly struggle with fundamental aspects of programming, such as the "basic sequential nature of program execution" [52]. This struggle with fundamental aspects prevents novices from employing advanced mechanisms. One such advanced mechanism might be tools for inspecting the run-time behavior, such as debuggers. A study on debugging behavior showed that novices used the debugger significantly less frequently than expert programmers, although they rely on progressive evaluation just as expert programmers do [24, 25, 35].

These differences between novice and expert programmers might also account for the lack of any results in the study described in Section 2.2.1 [27, 28]. As liveness implies new workflows, even expert programmers with a lack of *experience with liveness* should be considered novices in using liveness [33]. This, in turn, might explain the weak results of the studies reported in Section 2.2.2 and Section 2.2.3 [34, 65].

In summary, these results suggest that experience with liveness could be a relevant variable when investigating the effects of liveness.

### 2.3.2 Short Feedback Loops and Task Complexity

An assumption of live programming is that a short feedback loop (a short time span between making changes to source code and feedback on the change through seeing new dynamic behavior) generally benefits programmers.

However, the extensive body of research on system response times shows that the impact of short feedback loops is more nuanced [15]. Among other variables, the nature and *complexity of the task* moderate the effect of system response times [15, 20]. For example, studies show that a short response time can improve the performance on data entry tasks [41]. At the same time, for simple problem-solving tasks, a short response time decreases planfulness and worsens learning outcomes [60]. Finally, other preliminary results indicate that the same effects might not apply to complex problem-solving tasks [45].

This missing evidence for complex problem solving prevents the direct application of the results to software development, as, for example, debugging or program comprehension involves working on complex tasks. Further, many of the insights stem from research on user interface types, in particular on direct manipulation and command line interfaces. Thus, the applicability of the results to programming with its predominant text-editing interface might be limited even further.

Nevertheless, these insights about short response times, together with the weak results from existing studies on liveness, show that a short feedback loop is not necessarily beneficial by itself. At least the complexity or nature of the task has to be considered when we want to determine the usefulness of liveness.

### 2.4 Summary: Current State of Empirical Studies on Liveness

Overall, the results remain inconclusive due to the small number of studies that all focus on different variables. As live programming is becoming increasingly common in all kinds of contemporary programming systems, it becomes all the more relevant for developers of programming environments to know when and how liveness benefits programmers. A more refined and tested model of the factors influencing the impact of liveness might help them decide how to integrate liveness.

In particular, based on the future work of previous studies and the related research on the impact of task complexity, we argue that the existing studies of liveness were missing the aspect of task complexity.

### 2.4.1 The Impact of Liveness and Factors Influencing It

While there is no coherent theory describing the effects of liveness on programmers and the factors influencing them, the existing theories and studies provide a first glance at potentially relevant aspects. We briefly summarize the factors and potential effects that were covered by the theories and studies on liveness described above. The resulting collections of effects and factors are not a complete model yet but serve as a starting point to determine which aspects have been investigated. With enough insights from future studies, these collections might evolve into a coherent model in the future.

We speculate that liveness can have an impact on the following aspects:

**Program comprehension** Liveness may, for instance, influence the time to understand a program, the level of detail at which programmers understand it, programmers' confidence in their learnings [34], or make dynamic behavior easier to understand for novices [26, 27].

**Debugging** Liveness may influence debugging efficiency (time) or debugging effectiveness (number of solved defects) [47, 64].

**Program creation** Liveness may change how often programmers introduce defects when writing or changing code or reduce the time to repair defects as programmers notice wrong behavior earlier [34].

**Program design** Liveness may allow programmers to explore alternatives faster and thereby help them avoid local maxima in their program design [56, 57].

**Well-being** Liveness may reduce perceived emotional stress and lower mental strain while programming.

Further, we expect, among others, that the following factors influence the impact of liveness:

**Liveness level** To which degree is live feedback provided? Different levels provide different experiences of liveness. For example, updates after every edit action or only on explicit saving change the impression of immediacy of feedback [61].

**Response time** How fast is feedback made available? In particular, how much time elapses between completing a change and observing a change in the behavior resulting from that change [49]. Response times influence the impression of causality [30].

**Availability of introspection and intercession tools** Similar to the liveness level, the degree to which the tools allow programmers to get fine-grained insights into run-time state and behavior may influence the impact of the resulting experience of liveness.

**Overall skill level of programmers** The programming skill level influences tool usage, with expert programmers making more effective use of tools [25].

**Experience of programmers with liveness** Depending on the impact and the manifestation of liveness, programmers might require practice to benefit from it. Live features for making programming more accessible are typically usable without practice [26], but features designed to improve program comprehension through in-depth run-time information likely require prior experience.

**Task complexity** Task complexity might influence the degree to which programmers employ complex tooling. Results of a previous experiment on liveness suggest a relationship between task complexity and the effects of liveness [14, 65].

## 3 Experimental Design

To advance the progress towards a more complete view of the factors influencing the impact of liveness, we study the moderation effect [3] of task complexity on the effect of liveness through a controlled experiment.

### 3.1 Hypothesis

The main hypothesis for our experiment is: *"Task complexity moderates the impact of live tool support for introspection on debugging performance for programmers who are experienced with live programming tools."*

From our review of existing experimental evidence on liveness, we observed that previous experiments neglected two factors that may influence the impact of liveness: task complexity and participants' experience with liveness. Across the experiments, the complexity of the tasks was rather low, and participants had no experience with liveness of any kind.

Based on the described human factors studies on the relation between feedback loops and task complexity, we expect task complexity to substantially influence whether programmers can benefit from liveness [15, 20, 60].

Similarly, as live programming tools can profoundly impact the workflow of programmers [35], we expect that programmers need time to adapt their workflow to make use of it. As a result, their prior experience with liveness will determine whether they benefit from it.

Based on these observations, we investigated the moderation effects of task complexity given participants who have prior experience with liveness (for operationalization of factors, see Section 3.2). That is, we investigated whether liveness has different effects depending on the complexity of the task. We did not propose a specific direction of the moderation effect, as no strong prior results indicate how an increase in task complexity would influence the effects of liveness.

Does Task Complexity Moderate the Benefits of Liveness? A Controlled Experiment

We studied the impact of liveness on programming by investigating its effects on debugging since debugging has a clear goal that participants can work towards. Next to being suitable for experimental setups, debugging is also a relevant activity, as professional programmers spend considerable time on it [36, 43, 58].

We studied the effects of liveness from the perspective of exploratory programming environments, as this kind of liveness is also found in widely used programming environments, such as Chrome DevTools [5] and Jupyter notebooks [32]. The experience of liveness in exploratory programming environments results from dedicated tools for live introspection and intercession of the run-time state of a continuously running instance of the application to be developed.

## 3.2 Experiment Layout

We conducted the experiment as a within-subject study using a 2x2 factorial design. We decided on a within-subject study as we expected considerable differences in baseline debugging abilities between subjects (see Section 3.5).

The independent variables (IVs) were the presence of *live introspection tools* (IV1) and *task complexity* (IV2). The main dependent variable is *debugging performance* (DV1). To ensure that any effects we observed resulted from the availability of live introspection tools, we further regarded the *usage of live introspection tools* as another dependent variable (DV2). Participants' experience with liveness was set by recruiting students who all had the same exposure to an exploratory-style programming environment (see Section 3.5).

To reduce noise from the tasks, we aimed to let participants solve two tasks per condition, resulting in eight tasks.

In the following, we describe the operationalization of the independent and dependent variables.

### 3.2.1 Availability of Live Introspection Tools (IV1)

We took a broad perspective on exploratory programming environments as a form of live programming. In general, these programming environments provide means to work with a running instance of a program to be developed alongside the source code of the program. Therefore, they provide tools for introspection and intercession of run-time state and behavior. Smalltalk and Lisp systems are prime examples of such environments, as both fully support working with a live instance of the system to be developed. Contemporary environments provide a similar experience of exploratory programming for fields such as app development (Flutter tools [40]), web development (Chrome DevTools [5]), or scientific computing (Jupyter Notebooks [32]). They all incorporate some form of a running instance of the program and provide means to interact with it.

In this study, we used the Squeak/Smalltalk environment [29], as it offers a large variety of tools for exploratory-style live programming for software development in a general-purpose, object-oriented language. Many other environments with exploratory-style liveness features are for dedicated application domains (for example, Flutter tools or Chrome DevTools) or have special computation or execution models (for example, Jupyter notebooks or Excel). Each of these would require participants with a background either in the application domain or the computation or execution model. At the same time, the Squeak/Smalltalk tool set mostly consists of live tools that also exist in other environments, although in a different design (for instance the state inspector or the graphical meta-menu). Thus, we argue that any insights from Squeak/Smalltalk apply to other tools and environments with exploratory-style liveness. Finally, by using Squeak/Smalltalk we controlled for the participants' experience with the programming language and environment, as all participants have previously worked with the environment during two university courses (for details see Section 3.5).

We argue that the experience of liveness in exploratory programming environments [56, 57] is defined by two features. The first feature is the ability to *change the behavior of a running program*. When code is changed, the running program is updated automatically and behaves according to the updated code. The second feature is *full live introspection and intercession of run-time state*. Run-time state can be accessed from multiple entry points. The tools always reflect the current state in the system, and changes to state are directly applied. This also includes tool support for code evaluation in the context of the running system.

Based on this, we defined the following two conditions for the availability of live introspection tools.

**Control Condition (with)**   The base condition was an unrestricted Squeak/Smalltalk image that included hot-swapping of code, generic object inspectors and explorers that show the current state of an object, a graphical debugger that supports edit-and-continue, the graphical meta menu called "halo" to access run-time state from the user interface, and the ability to execute code in all tools.

**Experimental Condition (without)**   In the experimental condition, we prevented access to live introspection tools described above. The goal was to create a toolset that offers a programming experience similar to having a text editor with basic static features (syntax highlighting and checking, code navigation), an execution environment that needs to be invoked explicitly, and a debugger, such as the GNU Debugger (GDB).

We turned off hot-swapping, and thus, programmers had to restart the application for changes to take effect. Further, we disabled generic object inspectors and explorers, the halo meta-menu, and the ability to execute code in all tools. The internal logging mechanism (named `Transcript`) remained available.

As it is a common tool in all kinds of programming environments, the debugger also remained enabled. However, we restricted features related to live introspection. In particular, the inspectors embedded into the debugger only showed shallow snapshots of objects. Further, programmers could not change code directly within the debugger.

### 3.2.2 Task Complexity (IV2)

Related work from HCI and results from previous studies suggest that task complexity influences the effects of liveness. Task complexity is a complex concept with a variety of definitions [39, 48]. For this work, we define task complexity as a property of the task itself, in contrast to task difficulty, which depends on the relationship between task performers and the task. To control task complexity, we define task complexity using a collection of factors typically used in program maintenance studies [48]. The collection distinguishes task complexity factors along the variation points that can be influenced by task designers (task description, system, infection chain, patch, tool environment, general considerations) and a set of general complexity-contributing dimensions [39]. We used this collection to decide which tasks should be equal or different with regard to task complexity.

To observe the impact of live introspection tools, we shaped the task complexity so that the tasks prompted the usage of these tools. We expected the tools to be the most useful in following the infection chain and evaluating the patch,[1] thus we wanted the task complexity to originate from the infection chain and in parts from the patch.[2]

For simple tasks, the defect should be easy to localize and should already provide hints as to how it can be repaired. For complex tasks, the location of the defect should be more difficult to find, and there should be minimal hints as to how it should be repaired. To achieve this, we used defects of commission for simple tasks and defects of omission for complex tasks [4, 18].

**Simple Tasks (simple)**  For simple tasks, the defects are wrong selectors or class names in the source code. Thus, a defect was a wrong token in the source code, and as the defect is thus materialized in the source code, it can be spotted during defect localization.

Further, as a patch for such a defect is a change of the selector or class name, patch generation only involves determining the correct identifier to be used.

**Complex Tasks (complex)**  For complex tasks, defects are missing method sends. Consequently, the defect is not materialized in the source code directly and can not be easily spotted. During defect localization of the tasks in this condition, participants must fully understand the infection to determine when the system state became inconsistent. Correspondingly, patch generation is more complex, requiring participants to determine the required message send and a suitable location in the code.

**General Task Considerations**  To keep the task complexity comparable between tasks of the same complexity level, we considered various other factors when designing the tasks (for details on the task creation process, see Section 3.3; for an outline of the tasks, see Appendix C).

The task complexity should not have originated from the task description, system, tool environment, or general setup of the tasks, so the tasks should not exhibit, among others, overly long or inconsistent task descriptions, a convoluted system architecture, usage of obscure language features, or tight time limits (for an in-depth discussion of the task complexity factors, see Appendix D).

---

[1] We distinguish between the terms *defect*, which is the wrong section in the source code, the resulting *infection* in the system state, and the observable *failure* when the system does not behave as intended [66].

[2] A more in-depth discussion of the task complexity of the tasks of this experiment has been presented in prior work [48] (for an abbreviated version see Appendix D).

Concerning the infection chain, we aimed to keep the variance of the size and ambiguity of the infection chain comparable. Therefore, we used the same kind of failure for all tasks, which is behavior that runs but is incorrect with regard to correctly specified requirements [6]. So, the program does not crash but misbehaves, and we tell participants that the specifications are correct. We did not use crashes, because they provide an obvious starting point to investigate the system and potentially directly reveal the infection in the system state.

Also, all tasks only included a single defect within the application code. We explicitly told participants that there was only one defect and that the defect should not be found in the system code.

With regard to the patch participants needed to create, we also wanted to prompt the usage of live inspection tools. At the same time, we did not want participants to invest too much effort into determining the target behavior implemented by the patch. Thus, the tasks should make it obvious what the patch should do, but not necessarily how it can be done. Therefore, all tasks only require small changes, but in the case of simple tasks, only a wrong identifier needs to be changed, while for the complex tasks, participants need to write the missing message send. Further, to prompt participants to evaluate the path using live introspection, we did not provide tests.

### 3.2.3 Debugging Performance (DV1)

We measured debugging performance through *debbuging efficiency* as the time until a defect was repaired and *debugging effectiveness* as the number of correctly repaired defects.

The time to repair a defect began after we read the task to the participants and stopped when they proclaimed that the defect was repaired. This corresponds to *debugging efficiency* as used in other studies [12, 14, 46]. To ensure that we could collect enough data within an experiment session, we limited the task duration to 60 minutes. We expected participants to take longer than that only for complex tasks and potentially when participants did not have live introspection tools available. Thus, by limiting the maximum time to 60 minutes, we lower the measured times on these conditions, which puts these conditions at an advantage over conditions with simple tasks and conditions in which the live introspection tools are available.

Participants might apply different criteria to decide when the defect is repaired, thus introducing undesired variance to the time mesaurements. To account for this, we told participants that completing the task corresponds to committing changes to a project and that they should decide correspondingly.

Further, as a measure of the *debugging effectiveness*, we counted the number of correctly repaired defects, which, as there is one defect per task, corresponds to the number of correctly solved tasks [12, 14, 46]. After the participants had completed the experiment, we determined whether the task was solved correctly by examining whether the resulting behavior corresponded to the desired behavior described in the task without introducing other wrong behavior (see Section 3.3). There were no ambiguous cases, so only one grader decided on the correctness.

### 3.2.4 Usage Frequency of Live Introspection Tools (DV2)

To ensure that any effects we observed result from a change in tool usage, we also measured the relative usage frequency of the tools that constitute the liveness in the environment. We, therefore, recorded the interactions with tools showing static information and with live introspection tools (see Section 3.2.1).

We expected a pronounced increase in the usage frequency of live introspection tools when they were available. In case we did not observe an increase, the participant only worked with static tools, and we could not attribute differences in debugging performance to the changed tool set.

We tracked the usage frequency of tools through a background interaction tracking system. The tracking system recorded interactions with all tool windows, interactive code evaluation, and the opening of the graphical meta menu. We calculated two metrics characterizing to which degree live introspection tools were used: click ratio and the interaction time ratio. The click ratio is the proportion of clicks in live introspection tools to all clicks in programming tools. The interaction time ratio is the proportion of the time spent in live introspection tools to all the time spent in programming tools. We determined the time spent in a tool based on interaction streaks with one specific tool. A streak is a sequence of interactions with the same tool. The duration of a streak is the time between the first and the last timestamp of an interaction with the tool. To account for reading times, we added 5 seconds to the duration of streak. If an interaction with another tool immediately followed the last interaction, we only added the difference between the two interactions.

### 3.2.5 Participants' Experience with Live Introspection Tools

We argue that, when looking at the impact of liveness on debugging, participants first need to learn how to leverage liveness. In the context of our experiment, to make full use of live introspection tools, users need to adopt new workflows, such as keeping the program running instead of restarting it for every investigation, using a debugger and live evaluation instead of printing run-time values, or using Halos to explore the connection between UI elements and run-time objects instead of code reading. Thus, to ensure that potentially missing effects did not result from missing prior experience, we recruited participants with prior experience with liveness (see Section 3.5).

## 3.3 Scenario

To reduce the effects of learning about the program and the program domain, we used a single program for all tasks. We chose games as the program domain, as participants were familiar with developing basic games in Squeak/Smalltalk (see Section 3.5). We expected this to reduce the variance resulting from participants being unfamiliar with a program domain and only learning about domain concepts while working on the tasks. At the same time, games combine a variety of concerns, such as event handling, state propagation, file I/O, rendering, and algorithms. This variety allowed us to define tasks covering different program parts without overlap.

The program we have selected is the game "Jump-O-Drom", developed by a group of undergraduate students in a course on software architecture (see Figure 2). It is a multiplayer jump-and-run game in which most rules can be changed, resulting in many different game modes. It includes the following features:

- configurable game modes,
- configurable physics,
- configurable player appearance and controls,
- extensible collision handling,
- level editor,
- temporary effects on players,
- abilities for players,
- sound, and
- custom widgets and menu classes.

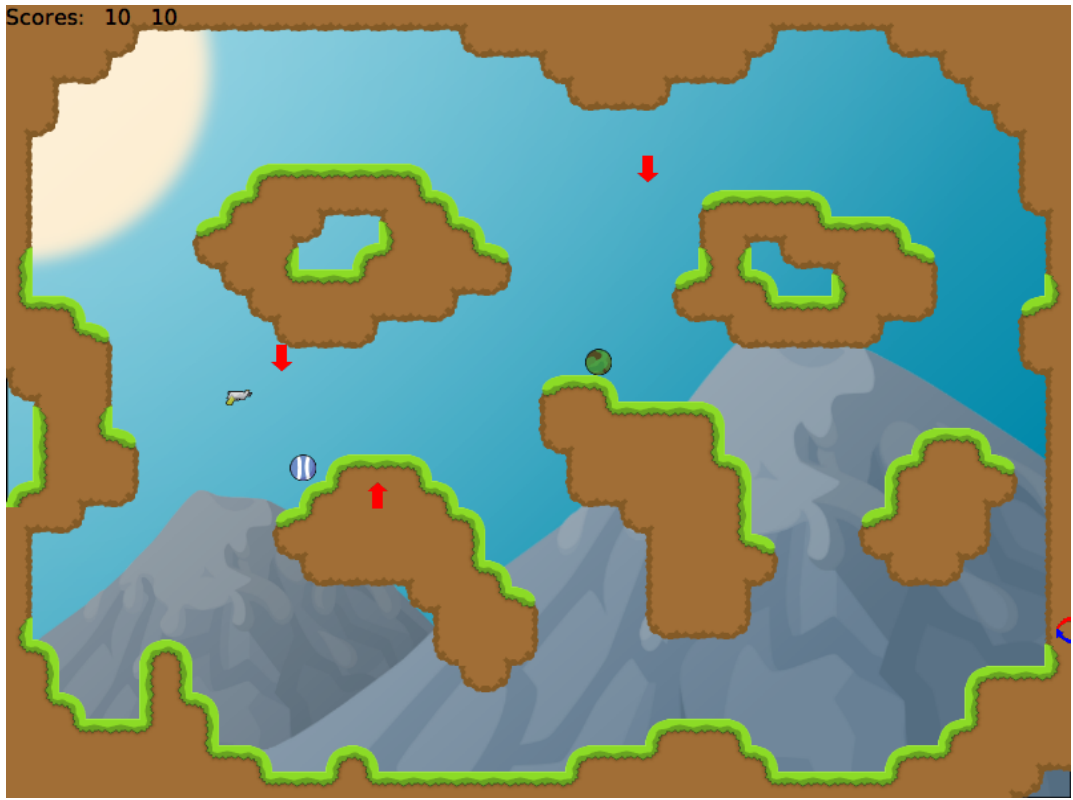

**Figure 2** A screenshot of the game Jump-O-Drom used as the program in our example study. The screenshot shows a running round. The blue-white and the green-brown circles are the players. In the currently active minigame, the goal is to reduce the score (top left) of the other player to zero. They can reduce the score of the other player by jumping on them.

Configurable features like game modes, physics, and player appearance and control are features that users can change through game settings. The extensible collision handling refers to the fact that the collision handling strategy of each entity can be changed at run-time.

With 3052 LOC the game has an average size for a game with the aforementioned features developed during the course (for more metrics, see Table 1). The size of the implementation does not limit task complexity, as even for small projects, complex debugging tasks can occur, depending on the number of code artifacts that need to be understood or the length of the infection chain. We removed any major idiomatic or architectural flaws to make the code base as approachable as possible, as we did not want participants to invest effort to navigate the source code.

**Table 1** Metrics characterizing the game Jump-O-Drom.

| Metric | | Metric | |
|---|---|---|---|
| #Packages | 8 | LOC | 3052 |
| #Classes | 73 | LOC / Method | 4.0 |
| #Methods | 759 | | |
| #Methods / Class | 10.4 | | |

### 3.4 Task Creation

Within each complexity level, we strove to create tasks with a similar level of complexity. However, the tasks inevitably varied slightly in their complexity. Thus, we planned to use two tasks per condition, resulting in eight tasks in total.

We created the tasks in several steps. First, we created candidate defects for both complexity levels. As we want to control task complexity, we decided to manually seed the defects instead of extracting defects from actual development activity, for example by reviewing the VSC history. We created 11 defects with wrong method selectors or class names and ten defects with missing method sends.

To ensure that the structural nature of the defects results in a lower or higher complexity of the debugging process as a whole, we cross-checked the resulting task complexity with an expert Smalltalk programmer who worked through the tasks. We recorded their debugging efficiency and observed their behavior. We then used the time measurements and the observations to discuss the challenges arising from each task. This led to the exclusion of three tasks whose perceived complexity did not match the intended complexity level.

To prevent learning effects between tasks, all eight tasks should affect a different concern and require participants to understand different artifacts in the game. We, therefore, determined the classes involved and the concerns of the game for each task. Based on this information, we determined the final eight tasks (see Appendix C).

Finally, we checked whether the eight tasks could be completed within the four hours of the experiment by asking two intermediate Smalltalk programmers to work on them and recording their time to complete each task.

### 3.5 Participants

We conducted the experiment with software engineering students enrolled at the Digital Engineering faculty of the University of Potsdam. In general, having students as participants in experiments impedes the generalization of results to professional programmers. However, with professional programmers as participants, we could not accurately control for previous experience with liveness. Also, to train professional programmers adequately on liveness to show a similarly high level of experience with liveness would be difficult to achieve in a reasonable amount of time [33, 59].

We recruited a total of 45 participants at two points in time over the course of two years. Students attended two consecutive compulsory courses during which they extensively worked on projects in the Squeak/Smalltalk live programming environment [21, 29]. We recruited students directly after they completed those courses. Consequently, all participants have had at least nine months of exposure to liveness and had a similar time between finishing the course and participating in the experiment.

All participants were reimbursed. To ensure no conflicts of interest, we only started recruitment and experiment runs after the grading of the university course was completed and final.

**Demographics of Participants**   We recorded the runs and questionnaires of 37 participants.[3] Due to a data loss, we lost the original demographics answers for 29 of the participants. Participants were asked to fill out the questionnaire again approximately three months after they answered it the first time. Three participants did not fill out the questionnaire a second time.

Participants were between 19 and 26 years old (median age: 21 years) when they participated. Nine participants identified as female, 24 as male, and one participant preferred not to answer.

We also asked participants to assess their experience with different programming aspects relevant to the experiment (see Figure 3). The data loss also included responses for the self-assessment. As participants repeated the self-assessment at a later point in time, they should be considered unreliable. Notably, participants rated their experience with live programming as low. We presume that this low rating results from participants not knowing the term live programming, as there are few explicit references to the term in lectures. An indication is that they rated their experience with the live environment Squeak as rather high at the same time.

To assess programming skill further, we used a skill test [19, 31, 55]. The skill test was based on 17 small programming exercises covering various language and standard library features, assuming that more experienced programmers are familiar with more of them than less experienced programmers [55]. The data loss did not affect the results of this skill assessment. Still, we needed to exclude the skill test of one of the 37 participants as they did not complete the test. Participants' scores were spread considerably, with a minimum score of 4 points (of 17 possible points) and a maximum score of 16 points (median score: 12). Overall, the skill level of participants is high enough for them to not count as novices anymore. Therefore, we expect them not to struggle with fundamental aspects of the tasks.

---

[3] We recruited 45 participants, but eight runs did not yield enough data to be included in the analysis.

**Does Task Complexity Moderate the Benefits of Liveness? A Controlled Experiment**

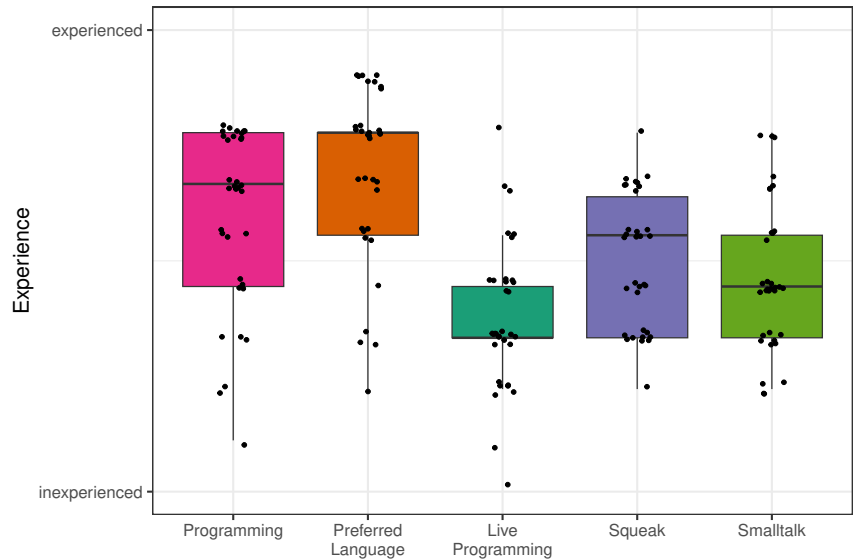

**Figure 3** A combined box and scatter plot showing participants self-assessed experience levels regarding: general programming, their preferred programming language, live programming, Smalltalk, and Squeak. Participants were asked to rate themselves on a 10-point scale from inexperienced to experienced.

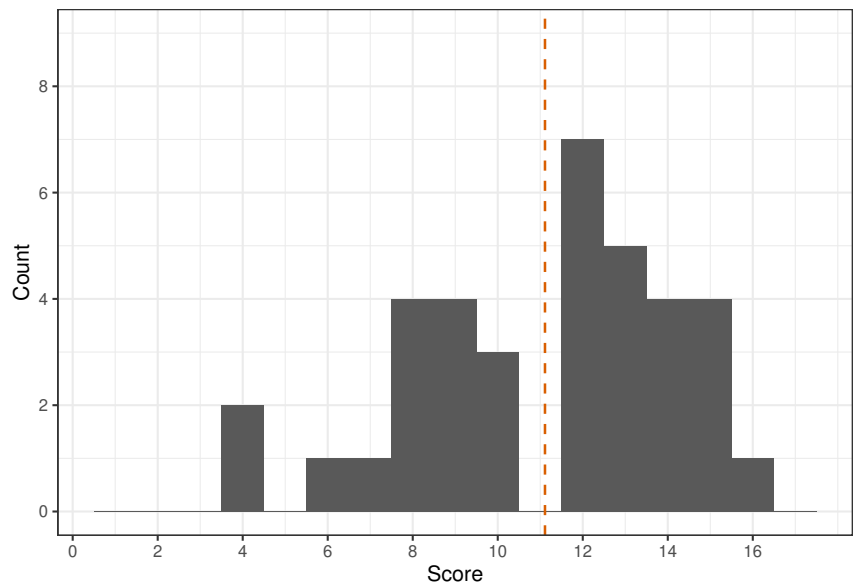

**Figure 4** A histogram of participants' scores in the skill test (red dashed line denotes mean).

### 3.6 Balancing of Conditions

As we employed a 2x2 experiment layout, participants worked in four conditions. We counterbalanced the conditions to account for the ordering effects of the within-subject design. A full counterbalancing would have resulted in 24 orderings, which we could not have assigned equally to participants, as we expected less than 40 participants. As a compromise, we applied a Latin square counterbalancing using a balanced Latin square [9]. We also balanced the assignment of tasks to conditions, resulting in 16 different configurations, which we assigned randomly to participants.

### 3.7 Experiment Procedure

We conducted all runs as remote meetings via Zoom. Participants received a pre-configured Squeak/Smalltalk image based on Squeak 5.3. Participants were allowed to change basic editing settings, such as whether lines are automatically indented, but were not allowed to load additional tools or packages.

We wanted to ensure that we gathered enough data from each run. Therefore, we first ran through all four conditions with one task for each condition, and after a break, we ran through the four again with another set of four tasks. With the 60-minute time limit, we ensured that even when participants took much longer than anticipated to complete two tasks, we still at least collected data for each of the four conditions.

For the most part, participants were guided through the experiment via a tool in the pre-configured programming environment (for a complete schedule, see Appendix B). Nevertheless, we explicitly introduced each new task to participants to ensure that participants read the instructions and successfully reproduced the failure.

To reduce learning effects, we introduced the relevant features of the game to participants, walked them through the main classes in the code, and recapped available programming tools. For all three steps, participants already used the environment, and we only instructed them on what to do to allow participants to adapt to the experiment setting. Finally, participants worked on a simple warmup task in the without condition. This warmup task should allow them to adjust to the missing tools in the without condition and familiarize them with the nature of the tasks. The complete introductory phase took between 30 and 60 minutes.

## 4 Results

The following analyses are based on the results of the 37 valid runs that completed at least four tasks (for results, see Figure 5 and Figure 6). Most participants did not complete all eight tasks, so we limited our analysis to only the first four tasks and ignored the second four tasks for all participants. The first four tasks were the same for all participants.

We first conducted the analysis of the moderation and main effect. To further understand the results, we conducted exploratory, post-hoc tests on the correlation of the debugging efficiency with interactions with dynamic tools.

**Does Task Complexity Moderate the Benefits of Liveness? A Controlled Experiment**

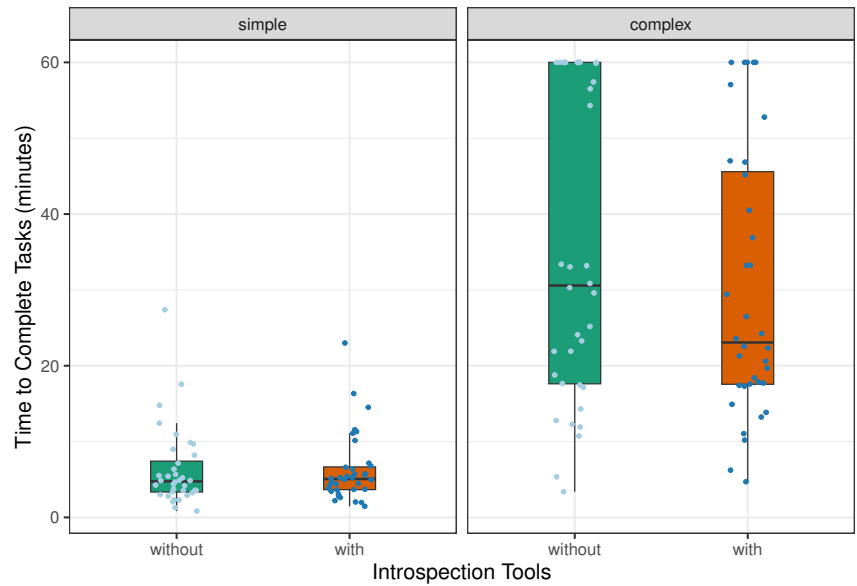

**Figure 5** A combined box and scatter plot showing the absolute times to complete the tasks grouped by task complexity and the availability of live introspection tools.

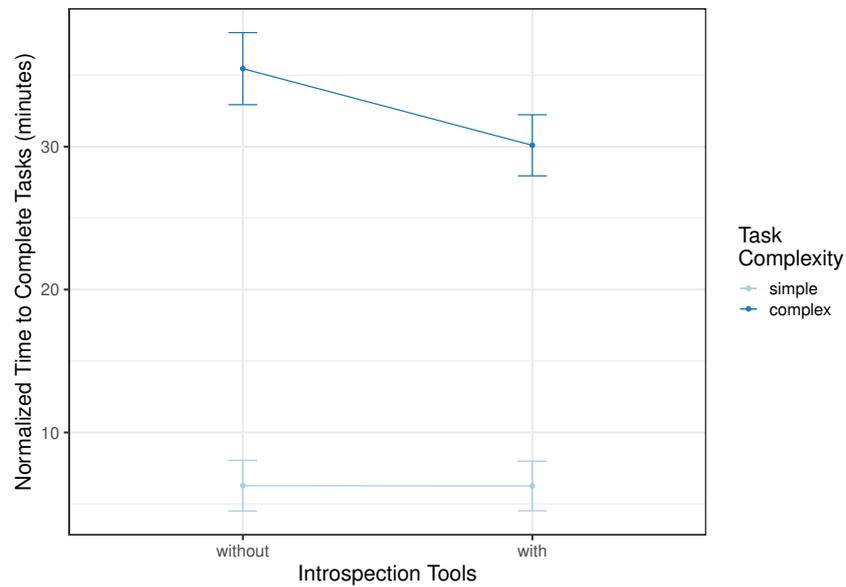

**Figure 6** Line charts showing the normalized means and error bars of the time to complete tasks when live introspection tools are available and when they are not distinguished between simple and complex tasks. The means and error bars are normalized per participant and condition using the Cousineau–Morey approach [2].

### 4.1 Analysis of Interaction and Main Effect on Debugging Efficiency

We tested the main hypothesis that *task complexity moderates the impact of tool support for live introspection on debugging performance for programmers experienced with live programming tools*. Due to the standard factorial repeated measures experiment layout, we used a standard two-way repeated measures ANOVA with a significance level of 0.05.

**Assumptions**   The assumption of sphericity was met, as both IVs only have two levels.

The assumption of no outliers was not met, as we had nine outliers in the 148 measurements. Four of these nine outliers were extreme outliers. We decided to keep the outliers in the data, as they are valid measurements of participants struggling with one particular task. We considered transforming measurements, but only four of the nine outliers could be resolved through transformations. As the transformations also complicate the interpretation of results, we decided against transformations.

The assumption of normally distributed data was also not met. The groups were not normally distributed (p < 0.01) according to the Shapiro-Wilk test of normality on the studentized residuals. However, in general, repeated measures ANOVAs are considered robust against non-normal data when sample sizes are above 30 (some simulations even suggest limits for the sample sizes as small as 10 [8]). Thus, we continued without transforming the measurements.[4]

**Result of ANOVA**   There was no statistically significant interaction between task complexity and the availability of live introspection tools on debugging efficiency ($F(1, 36) = 1.84, p = 0.18$). Thus, our initial hypothesis was invalidated.

The main effect of the availability of live introspection tools showed a statistically significant difference in participants' debugging performance ($F(1, 36) = 4.54, p = 0.04$). Thus, in the setting of our experiment, participants benefited from having live introspection tools available.

The main effect of task complexity showed a statistically significant difference in the debugging performance of the participants ($F(1, 36) = 82.95, p < 0.001$). Thus, the task design resulted in tasks with significantly different task difficulty, suggesting that the tasks were of different complexity.

### 4.2 Effect on Debugging Effectiveness

We did not analyze the effect of live introspection tools on debugging effectiveness, as there is not enough variance in the correctness of the submitted patches. During the experiment runs, we observed that the setup did not lead to wrong patches and thus decided against an in-depth analysis. The counts confirm this; only 2 of the

---

[4] We checked transformations after the initial test, but neither the square root nor the log transformation resulted in normally distributed data. We refrained from more complex transformations to keep the results easier to interpret.

**Does Task Complexity Moderate the Benefits of Liveness? A Controlled Experiment**

148 patches were wrong, one for a simple task with live introspection tools and the other for a complex task without live introspection tools. For comparison, there were 15 timeouts, all for complex tasks, five with live introspection tools, and ten without.

We assume the missing variance in the correctness of the patches results from the experiment setup. By telling participants to work on a task until they are confident in their solution, we primed participants to solve the tasks correctly. Correspondingly, they would rather work until the timeout than submit a wrong patch.

### 4.3 Post-hoc Analyses of Usage of Live Introspection Tools

We conducted post-hoc analyses to check whether the availability of live introspection tools did affect the usage frequency of live introspection tools. Therefore, we conducted two-way repeated measures ANOVAs with a significance level of 0.05. We conducted ANOVAs on both the click and the time ratio, even though they depend on each other. We did so to ensure that any differences we might observe result from the actual usage frequency and not from a systematic bias in one of the metrics (see Figure 7). For both metrics, the usage frequency of live introspection tools is greater than zero in the without condition. This non-zero frequency results from the debugger being part of the group of live introspection tools, even when the live features are disabled in the without condition.

Again, the assumption of normality was not met for both metrics, but the sample size was large enough to continue. There were no significant outliers, and the assumption of sphericity is again met, as both independent variables were dichotomous.

Both ANOVAs show significant main effects and a significant interaction effect. We only report the results for the time ratio; the results for the click ratio are similar. As expected, the main effect of the availability of live introspection showed a statistically significant difference in the time ratio ($F(1, 36) = 48.23, p < 0.001$). Also, the main effect of task complexity showed a statistically significant difference ($F(1, 36) = 81.15, p < 0.001$). Notably, there is also a significant interaction between the availability of live introspection tools and the task complexity ($F(1, 35) = 4.558, p = 0.04$)[5] (see Figure 7).

## 5 Discussion

Based on the analysis, we discuss the conclusions we drew from the results. To support the interpretation of the results, we also discuss threats to the internal and external validity of the results. Based on these insights, we outline future work to determine the effects of liveness in programming environments.

---

[5] We ran the simple main effect analysis, and all simple main effects are significant.

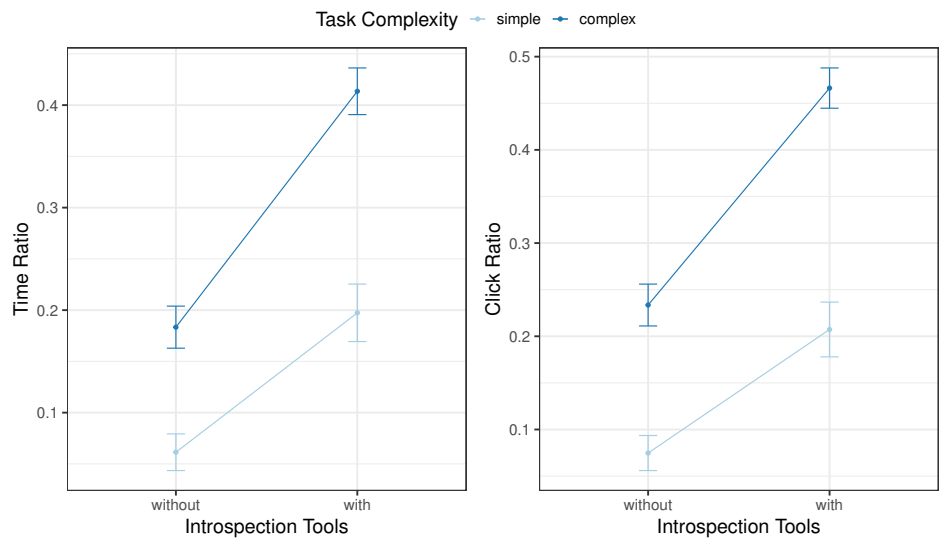

**Figure 7** Two line charts showing the mean time ratio and the click ratio when live introspection tools are available or not distinguished by the task complexity.

## 5.1 Results

The main hypothesis to be tested with this experiment was *"task complexity moderates the impact of tool support for live introspection on debugging performance for programmers experienced with live programming tools"*. The non-significant interaction between task complexity and the availability of live introspection tools invalidates this hypothesis.

However, the non-significant interaction might be the result of the sample size, as the interaction is significant at the significance level 0.2. In the charts a tendency towards an interaction can be seen that suggests that at a higher task complexity, the live introspection tools reduce the time needed to debug a task more than they do at a low task complexity (see Figure 6). While we determined the required sample size in a pre-study power analysis on pilot results, our sample size only satisfied the requirement for the main effect. The sample sizes for interactions typically need to be larger; thus, we underestimated the required sample size. Further, the results show a significant interaction between live tool availability and task complexity with regard to the ratio of time spent using the tools. The increase in live introspection tool usage between tasks without and tasks with the tools was steeper for complex tasks than for simple tasks (see Figure 7). Independent of the main results of the experiment, this observation suggests that future experiments should consider using tasks that tend to be complex, as they may increase participants' exposure to the tools.

Despite the invalidated main hypothesis, the main effect was statistically significant, so the availability of live introspection tools did improve participants' debugging efficiency. There was a significant difference in the time spent using the tools between tasks without and tasks with tools. Thus, we are confident that the availability of the tools was the main factor influencing participants' debugging efficiency.

Does Task Complexity Moderate the Benefits of Liveness? A Controlled Experiment

The statistically significant result of this experiment is in contrast to the inconclusive results of previous experiments. One explanation for the significant results might be the participants' prior experience with liveness. Another explanation might be the particular liveness experience tested in this experiment. Whether these factors make a difference remains a question for future work.

**Relation to Results from Experiment on Continuous Visual Feedback**   One of the prior experiment on liveness in the Forms/3 environment suggested task complexity as a relevant moderator variable for future work [14, 65]. The authors of that experiment suggested task complexity because liveness had a different impact during the two tasks in that experiment: During the task that involved debugging the control logic of a seven-segment display, participants did not benefit from the liveness features, but when they worked on the task that only involved arithmetic operations, they did benefit from liveness.

Their results match our observations in that they suggest task complexity influences liveness's effect. When looking at their experiment setup using our notion of task complexity, their task on the seven-segment display had a lower complexity than the task involving arithmetic operations. The seven-segment display task was graphical and had almost no infection chains, as the erroneous graphical state was displayed directly next to the spreadsheet cell computing the state [65, therein Figure 3]. Further, as each cell could be checked in isolation, the defects had a simple control flow, as most cells only consisted of a single conditional expression. In contrast, the arithmetic task had a higher complexity. Participants could not easily check the results of cells computing sub-expressions of the arithmetic formula. Instead, they would have needed to trace the erroneous computation from the cells showing the final result. For one defect, participants would even need to write the expression for an empty cell.

The tasks in our experiment are more complex than the two tasks in that prior experiment. Thus, the result of our significant overall effect might stem from having tasks that are complex enough. Further, while not significant, the effect of liveness seems higher with complex tasks, paralleling the observations from the prior experiment. Additionally, we argue that our expeiriment's results have a higher internal validity, as our participants had sufficient exposure to the environment beforehand, while none of the participants in the Forms/3 experiment had used the environment.

### 5.2  Threats to Validity

We identified two main and one minor threat to the internal validity of our results.

First, we removed tools that participants were used to. Typically, experiments with programming tools add a new tool in the experimental condition. In this experiment, however, we removed tools in the experimental condition. As participants had prior experience with Squeak/Smalltalk, we removed tools they were potentially used to. Thus, the effect we observed might partly result from participants working without their familiar tools. We counteracted this effect by letting participants work on the warmup task without live introspection tools, thereby giving them time to adjust to the reduced tool set. Further, while they used Squeak/Smalltalk for two semesters, they only used it in two of typically ten courses and were exposed to several other languages whose toolsets resemble the without condition. Thus, we can assume that participants could adjust to the new setting. Still, we can not exclude that the removal of tools itself had a considerable influence.

The second threat to internal validity is the selection of tools that we deactivated. Specifically, the deactivation of live object inspectors might have biased results. Without the live object explorers, participants could still explore objects in the debugger but they could only see the print strings of objects referenced from local or instance variables. They could not inspect the referenced objects further without stepping to a corresponding method. This additional step can lengthen the time needed to access information. We argue that the graphical, interactive, and live object explorers are part of the live introspection tools, and thus, we needed to deactivate them. However, whether the additional effort was so high that it dominated the effects of the complete live introspection toolset should be clarified in future experiments.

A minor threat to the internal validity is the lost data from our demographics and self-assessment questionnaires. As the majority of the participants filled out the questionnaire at a later point in time, the results were unreliable. At the same time, the resulting characterization of the participants is not essential to the experiment's result but illustrates the participants' general characteristics.

We identified the task design as the main threat to the generalizability of the results. We constructed the tasks to prompt the usage of dynamic tools, including live introspection tools. In practice, programmers might not use dynamic tools as often during debugging as they did when working on our tasks. Thus, any effect of live introspection tools in practice will likely be lower than what we observed. Further, while there is a notable difference in task complexity between the simple and complex tasks, the complex tasks are not yet representative of complex debugging tasks in practice [17]. For instance, our complex tasks have comparatively short infection chains, can be observed consistently, and can be examined using ordinary debugging tools. We did not design our tasks to be representative of very complex debugging tasks but designed them to exhibit a major difference in task complexity while remaining solvable in the given timeframe. Still, when programmers work on tasks that are more complex, they might use very different strategies, thereby either increasing or decreasing the impact of live introspection tools.

### 5.3 Applicability to Different Forms of Liveness

Our experiment yielded insights on the effect of liveness, as seen from the exploratory programming perspective. We argue that the results directly apply to other environments that correspond to the exploratory programming perspective and offer a comparable set of tools for live introspection.

By taking a more general, and yet speculative perspective on the mechanisms underlying our observation, the results might also apply to tools that correspond to the live programming and live coding perspectives. Through liveness, programmers can get immediate access to dynamic information. Thereby, liveness features reduce the cost of accessing dynamic information in terms of time and mental effort programmers must invest. Thus, programmers might use dynamic information more often, and thus more often in situations in which it is the most useful information to solve a problem [54] and thereby become more efficient overall.

When we assume that the immediate access to dynamic information is the main cause for the improved debugging efficiency, then our results would directly apply to live programming tools. Live programming tools focus on getting dynamic information to programmers automatically by re-executing the program on changes and updating visualizations. The reduction in costs of accessing dynamic information might thus lead to a similar improvement in debugging efficiency.

Our results do not directly apply to live coding tools, as the live coding perspective is characterized more by the goal of controlling a real-time effect than the tools' feature sets. Some of these tools include features that resemble those of live or exploratory programming tools, in which case our results may apply. Some features are specifically designed to make real-time process control possible or easier. Thus, at least in debugging or program comprehension scenarios, they will not improve access to dynamic information.

Whether the improved debugging efficiency is the result of the decreased cost of accessing dynamic information or whether it results from specific features still needs to be settled by future studies.

### 5.4 Future Work

This experiment has shown one setting in which liveness was beneficial. However, the question of how and why liveness helps programmers remains open. Based on our observations, we suggest some specific factors for future studies.

**Short Access Time or Introspection Tools**   The experience of liveness based on live introspection tools stems from two aspects: the features of the introspection tools themselves and immediate access to information when the need arises. For instance, the defining feature of the Halo tool is that it allows programmers to navigate from UI elements to underlying objects. The fact that programmers can use it to get objects from within a running program makes it even more useful, as they can quickly get information on these objects when the need for it arises. The question is whether the Halo tool would still be useful when immediate access is not possible anymore, for instance, because it only works on a post-mortem state snapshot from a program run for which programmers manually added instrumentation to the source code.

**Background of Participants**   One primary difference between this experiment and previous ones is the considerable exposure of participants to liveness before the experiment. At the same time, when they participated, the participants were junior programmers with regard to their overall programming experience. Repeating a similar experiment with more experienced programmers who, in addition, had considerable exposure to liveness may help distinguish between the effects of general programming experience and the specific experience with liveness.

**Towards a Theory of the Impact of Liveness**   Existing theories on liveness can guide creators of programming systems in creating a liveness experience, but they do not strive to explain why and how liveness helps programmers in detail. In this article, we reviewed the factors that related studies assumed to influence the impact of liveness. However, at this stage, the result of the review is merely a collection of potential impacts and factors. To make considerable progress in understanding the impact of liveness, we need a theory explaining the mechanisms of using liveness. Whether this theory should be at the cognitive level or only the process level is an open question.

## 6  Conclusion

While liveness is generally regarded desirable, its specific effects on programmers have not been studied extensively. A few experiments and studies on liveness exist, but they remain inconclusive.

We reviewed these experiments and studies to create a first collection of potential effects of liveness and moderating factors. We hope this collection helps future researchers study liveness or evaluate live programming features.

Based on our collections of effects and factors, we set up an experiment to test the hypothesis that task complexity moderates the impact of live introspection tools on the debugging performance of participants with prior experience with the liveness manifested in these tools. In the analysis of the results, we found no significant moderation effect, which might be attributable to the small sample size, as the measurements show a visible trend toward an interaction between task complexity and debugging efficiency. At the same time, we found that live introspection tools significantly improved debugging efficiency.

The results of our experiment are only a step towards understanding how liveness takes effect. For researchers interested in liveness, the results suggest that task complexity and participants' experience with the kind of liveness to be studied are essential factors. Further, for designers of programming environments, the results show that programmers can benefit from live introspection tools.

**Data Availability**   The full experiment data is available on Zenodo, including the experiment setup (environment setup, system, task descriptions, procedures, questionnaire), results (measurements per task, event sequences, questionnaire results), and the statistical analysis scripts [50].

**Acknowledgements**   We sincerely thank the anonymous reviewers for their detailed and valuable feedback. This work was supported by Deutsche Forschungsgemeinschaft (DFG) grant #449591262 and the HPI–MIT "Designing for Sustainability" research program.[6]

---

[6] https://hpi.de/en/research/cooperations-partners/research-program-designing-for-sustainability.html (accessed October 25, 2024).

**Does Task Complexity Moderate the Benefits of Liveness? A Controlled Experiment**

# A  R Configuration

R version 4.4.0 (2024-04-24)
Platform: x86_64-pc-linux-gnu
Running under: Ubuntu 22.04.4 LTS

Matrix products: default
BLAS:   /usr/lib/x86_64-linux-gnu/blas/libblas.so.3.10.0
LAPACK: /usr/lib/x86_64-linux-gnu/lapack/liblapack.so.3.10.0

time zone: Europe/Berlin
tzcode source: system (glibc)

attached base packages:
stats;  graphics; grDevices utils;  datasets; methods;  base

other attached packages:
viridis_0.6.5; viridisLite_0.4.2; moments_0.14.1;  scales_1.3.0; RColorBrewer_1.1-3;
likert_1.3.5; xtable_1.8-4; rstatix_0.7.2; ggpubr_0.6.0; lubridate_1.9.3; forcats_1.0.0;
stringr_1.5.1; dplyr_1.1.4; purrr_1.0.2; readr_2.1.5; tidyr_1.3.1; tibble_3.2.1;
ggplot2_3.5.1; tidyverse_2.0.0; reshape2_1.4.4

loaded via a namespace (and not attached):
utf8_1.2.4; generics_0.1.3; stringi_1.8.3; lattice_0.22-5; hms_1.1.3; magrittr_2.0.3;
grid_4.4.0; timechange_0.3.0; plyr_1.8.9; backports_1.4.1; gridExtra_2.3; fansi_1.0.6;
abind_1.4-5; mnormt_2.1.1; cli_3.6.2; rlang_1.1.3; munsell_0.5.1; withr_3.0.0;
tools_4.4.0; parallel_4.4.0; tzdb_0.4.0; ggsignif_0.6.4; colorspace_2.1-0; broom_1.0.5;
vctrs_0.6.5; R6_2.5.1; lifecycle_1.0.4; car_3.1-2; psych_2.4.3; pkgconfig_2.0.3;
pillar_1.9.0; gtable_0.3.5; glue_1.7.0; Rcpp_1.0.12; tidyselect_1.2.1; rstudioapi_0.16.0;
nlme_3.1-163; carData_3.0-5; compiler_4.4.0

# B  Experiment and Task Schedule

## B.1  Experiment Schedule

The schedule for one run was:

1. introduction, welcome, study agreement
2. reimbursement
3. introduction to the gameplay of the game
4. overview on the packages making up the game
5. introduction to the two modes in the environment:
   a. mode 1: recap of available dynamic tools
   b. mode 2: illustration of limitations

6. introduction of task management tool
7. warmup task to familiarize participants with the without condition and the nature of the tasks
8. tasks 1 - 4
9. tasks 5 - 8
10. debriefing interviews: demographics, skill self-assessment, skill test

### B.2 Task Schedule

The schedule for each task was:

1. We announce that a new task begins.
2. Participants start the task via the task management tool.
3. The tool changes the condition if necessary and loads the defect.
4. We read the instructions to reproduce the failure and the description of the wrong behavior and the expected behavior to the participants. We observe whether they successfully reproduce the failure.
5. On our signal, they start the time measurement in the task management tool.
6. Participants work on the task and press a button in the task management tool when they decided that they completed the task.
7. They notify us that they completed the task, and we start preparing the next task with them.

**Does Task Complexity Moderate the Benefits of Liveness? A Controlled Experiment**

## C  Tasks

**Table 2** The eight tasks used in the experiment. Simple tasks are denoted with (s) and complex tasks with (c). The statistical analysis includes the results of the the first two tasks of each complexity levels (in italics).

| task | defect | failure | type | concerns |
|---|---|---|---|---|
| *1 (s)* | Minigame uses wrong class to initialize collision strategy of Player objects. | Players always fall through the level. | wrong class | player collision handling |
| *2 (s)* | Swapped conditional selector in LevelEditor file handling | Level editor duplicates the .json file ending for level files. | wrong selector | serialization |
| 3 (s) | Projectile collision strategy accesses projectile instead of removing it. | Destruction projectiles destroy all blocks in their way. Projectiles should only destroy one block and then disappear. | wrong selector | block collision handling |
| 4 (s) | Swapped conditional selector in LevelEditor UI rendering | UI indicates that the grid is enabled, but blocks can be placed freely. | wrong selector | level editor user interface |
| *5 (c)* | Updating player scores uses Player instance instead of numeric id in dictionary access. | The score counter shows four scores for a level with two players. A defeated player is instantly removed from the game even when only hit once. | missing send | game life cycle |
| *6 (c)* | Missing method send of removal method when editing settings | The items setting can not be deactivated anymore. | missing send | settings |
| 7 (c) | Missing method send to Player instances to collect the initial ability missing in PlayerSpawner | Initial abilities can not be used. | missing send | player life cycle |
| 8 (c) | Missing method send of reset method in HotBombBomb minigame | The bomb is directly passed to the next player without any delay and the game ends. | missing send | mini game life cycle |

## D  Discussion of Task Complexity

The following is an abbreviated version of a more in-depth analysis of the task complexity in this experiment based on the exemplary analysis of one task in prior work [48].

### D.1  System

The selected system determines the way we can influence subsequent variation points. We wanted to investigate the usage of dynamic tools in a scenario beyond a small module, but needed to stay within realistic time limits. We chose a small game as the system, as games combine a variety of concerns such as event handling, state propagation, file I/O, rendering, and algorithms.

Regarding *quantity (input)*, the game has a large feature set and the source code can be considered small (for details, see Section 3.3).

We were mostly interested in how programmers apply dynamic tools to fix a defect, not how they use them to learn about a system in general. Thus, we provided *guidance (input)* with regard to the general system behavior by introducing the module structure and important classes of the game. We used a fixed script to avoid providing any additional guidance that might influence the complexity of the individual tasks.

Similarly, as we were not interested in how programmers explore the present behavior of a system, we made the description of the game behavior *redundant (input)* by giving an interactive tutorial on the gameplay.

Concerning *clarity (input)*, we wanted to ensure that we do not observe tool usage resulting from unnecessarily complex code or a convoluted architecture. Thus, we ensured the project contained no significant idiomatic or architectural flaws.

### D.2  Task Description

We were not interested in how or how well participants could comprehend the description of the failure. Thus, we aimed to reduce the complexity of the task description.

For one, we aimed at reducing the complexity by keeping the *quantity (input)* down with a concise description of the task. We included the steps to reproduce the failure, a description of the observable symptoms, and a description of the expected behavior. Further, we aimed to keep the description *clear (input)*. Therefore, we used a consistent structure throughout all tasks, which distinguishes between the steps to reproduce the failure and the observable as well as the desired behavior. The structure was visually reinforced through a dedicated graphical tool presenting the tasks. Further, we used consistent vocabulary for interactions, parts of the game, and observable behavior throughout all tasks.

We wanted participants to make use of the tools to generate and test hypotheses about the failure. Thus, while the task description should be clear, it should at the same time only provide little *guidance* about the actual process of repairing the failure. Therefore, we aimed to give as few hints on the source code location of the defect as possible. For example, to not give away the class in which the defect is located, we did not use any terms related to the class name.

### D.3 Infection Chain

Our main interest in this study was how participants use dynamic tools to determine the defect location. Thus, we needed to make the defects complex enough for participants to spend considerable effort on locating them, while at the same time keeping them doable in the available time.

We used the *size (output)* of the infection chain to ensure that the defects could be found in the available time frame. Therefore, we spread the defect, infection propagation, and failure among few classes. Also, we did not want to observe special debugging techniques, thus we limited the tasks to one defect per failure.

At the same time, we needed to ensure that participants had to invest enough effort into locating the defect so that they had a reason to use dynamic tools. To achieve that, we influenced the *clarity (output)* of the infection chain by using wrong behavior resulting from a programming error. Finally, to distinguish between different levels of complexity, we used defects of commission for simple tasks, and defects of omission for complex tasks (for details, see Section 3.2.2).

### D.4 Patch

For our study goal, the complexity of the patch should ideally trigger the usage of dynamic tools to determine suitable source locations, inspect objects, and evaluate potential solutions. At the same time, to get comparable observations, we wanted to keep determining the target behavior simple. So, determining what to implement should be simple, while determining how to implement it should be complex.

The *clarity (output)* of the location and code for the patch differs between defects of omission and commission (for details, see Section 3.2.2).

Determining what to implement is kept simple due to the limited number of *redundant (output)* target behaviors. Due to the small size of the defects, participants should be able to describe the target behavior given that they correctly identified the defect. To make deciding on a target behavior even simpler, we aimed to prevent *conflicting goals (output)* by explicitly asking participants to work on the patch until they are as sure of it as they would when committing it to one of their own projects.

### D.5 Tool Environment

Regarding the *clarity (process)* of using the tools to observe the behavior, we did not provide automatic tests, as they would be obvious starting points. Further, concerning the *quantity of paths (process)* to employ the tools, we provided no hints on what they should use. To make sure participants are aware of all potential tools, we briefly recap the available tools at the beginning of a run. Finally, to prevent that a larger *quantity of steps (paths)* to get to relevant information prevents the usage of tools, we also recapped keyboard shortcuts and context menus.

## About the authors

**Patrick Rein** is a member of the Software Architecture Group of the Hasso Plattner Institute at the University of Potsdam. He is working on improving the usability of programming tools so programmers can use the tools most useful to them. His current research interests include the design and impact of live and exploratory programming systems. Contact Patrick at patrick.rein@hpi.uni-potsdam.de.
https://orcid.org/0000-0001-9454-8381

**Stefan Ramson** is a member of the Software Architecture Group of the Hasso Plattner Institute at the University of Potsdam. He regards the design of programming systems as the intersection of notation, interface design, psychology, and ergonomics. His current research interests include live and exploratory programming systems, alternative input methods, visual languages, and natural programming. Contact Stefan at stefan.ramson@hpi.uni-potsdam.de.
https://orcid.org/0000-0002-0913-1264

**Tom Beckmann** is a member of the Software Architecture Group of the Hasso Plattner Institute at the University of Potsdam. He is working on structured editing for general-purpose languages to better support the integration of tools. His current research interests include programming tool design, as well as editing and input methods for programming. Contact Tom at tom.beckmann@hpi.uni-potsdam.de.
https://orcid.org/0000-0003-0015-1717

**Robert Hirschfeld** leads the Software Architecture Group at the Hasso Plattner Institute at the University of Potsdam. His research interests include dynamic programming languages, development tools, and runtime environments to make live, exploratory programming more approachable. Contact Robert at robert.hirschfeld@hpi.uni-potsdam.de.
https://orcid.org/0000-0002-4249-6003